\documentclass{article}
\usepackage[T1]{fontenc} 
\usepackage[utf8]{inputenc} 
\usepackage{ismir,amsmath,cite,url}
\usepackage{graphicx}
\usepackage{subcaption}
\usepackage{color}
\usepackage[export]{adjustbox} 

\usepackage{booktabs}
\usepackage{lineno}
\usepackage[bookmarks=false]{hyperref}

\title{Automatic Estimation of Singing Voice Musical Dynamics}

\multauthor
{Jyoti Narang$^{\flat *\thanks{*These authors contributed equally to this work.}}$ \hspace{1cm} Nazif Can Tamer$^{\flat *\footnotemark}$ \hspace{1cm} Viviana De La Vega$^{\sharp}$ \hspace{1cm} Xavier Serra$^{\flat}$} 
{
  $^{\flat}$ Music Technology Group, Universitat Pompeu Fabra, Barcelona, Spain\\
  $^{\sharp}$ Escuela Superior de Música de Cataluña (ESMUC), Barcelona, Spain\\
  {\tt\small jyoti.narang@upf.edu, nazifcan.tamer@upf.edu,}\\
  {\tt\small vivianadelavega@gmail.com, xavier.serra@upf.edu}
}

\def\authorname{F. Author, S. Author, and T. Author}

\begin{document}

\maketitle

\begin{abstract}
Musical dynamics form a core part of expressive singing voice performances. However, automatic analysis of musical dynamics for singing voice has received limited attention partly due to the scarcity of suitable datasets and a lack of clear evaluation frameworks. To address this challenge, we propose a methodology for dataset curation. Employing the proposed methodology, we compile a dataset comprising 509 musical dynamics annotated singing voice  performances, aligned with 163 score files, leveraging state-of-the-art source separation and alignment techniques. The scores are sourced from the OpenScore Lieder corpus of romantic-era compositions, widely known for its wealth of expressive annotations. Utilizing the curated dataset, we train a multi-head attention based CNN model with varying window sizes to evaluate the effectiveness of estimating musical dynamics. We explored two distinct perceptually motivated input representations for the model training: log-Mel spectrum and bark-scale based features. For testing, we manually curate another dataset of 25 musical dynamics annotated performances in collaboration with a professional vocalist. We conclude through our experiments that bark-scale based features outperform log-Mel-features for the task of singing voice dynamics prediction. The dataset along with the code is shared publicly for further research on the topic.   

\end{abstract}
\section{Introduction}\label{sec:introduction}

Musical dynamics, such as \emph{piano} and \emph{forte}~\cite{patterson1974musical}, are key elements in adding expressiveness to the singing voice~\cite{fabian2014expressiveness}. They enhance overall performance and facilitate the conveyance of the desired emotional impact~\cite{miller1996art}. Despite extensive research on the singing voice, the analysis of dynamics in this context has received limited attention for several reasons.
Firstly, annotating dynamics is an expensive process that requires repeated listening to audio tracks to accurately identify the dynamics category. Secondly, unlike other musical features such as pitch or tempo, the categorization of dynamics is not clearly defined, and even the same annotator may interpret a piece differently on multiple listens. Finally, a significant challenge for modern deep learning applications is the lack of reliable, existing dynamics based annotated datasets that can be used for the development of automatic analysis systems~\cite{bous2023analysis}.

Despite the challenges of dynamics-based annotations for the singing voice, investigating dynamics in singing performances is worthwhile. On one hand, dynamics are a key component of expressivity in a music performance~\cite{sundberg1994perceptual, bishop2014performing}. On the other hand, dynamics are also an integral part of the music writing tradition~\cite{patterson1974musical, berndt2010modelling}. The use of dynamics in Western classical music evolved significantly from the Baroque period to the Romantic era. Particularly during the Romantic era, when expressivity became prominent, the annotation of dynamics alongside the score became widespread and accepted as part of the composition process. Composers frequently utilized symbols such as \textit{forte}, \textit{piano}, \textit{crescendo}, and \textit{diminuendo} to convey their desired variations in musical dynamics, and adhering to the dynamics instructions given by the composers became an important part of a Classical music performance.


While dynamics is a musical concept, its automatic estimation for music performance analysis relies on properties derived from audio signals. The audio characteristic most similar to musical dynamics is loudness or perceptual intensity. However, the mapping of musical dynamics to audio-based features from Music Information Retrieval (MIR) technologies is still not clearly understood. %
Extensive research exists on dynamics and tempo as expressive dimensions for Western classical piano performances~\cite{widmer2004computational}. However, unlike piano, there are almost no publicly available dynamics-based annotated datasets for the singing voice, which hinders the development of such technologies for the vocal performance analysis.
\begin{figure*}
        \centering
     \includegraphics[scale=0.53]{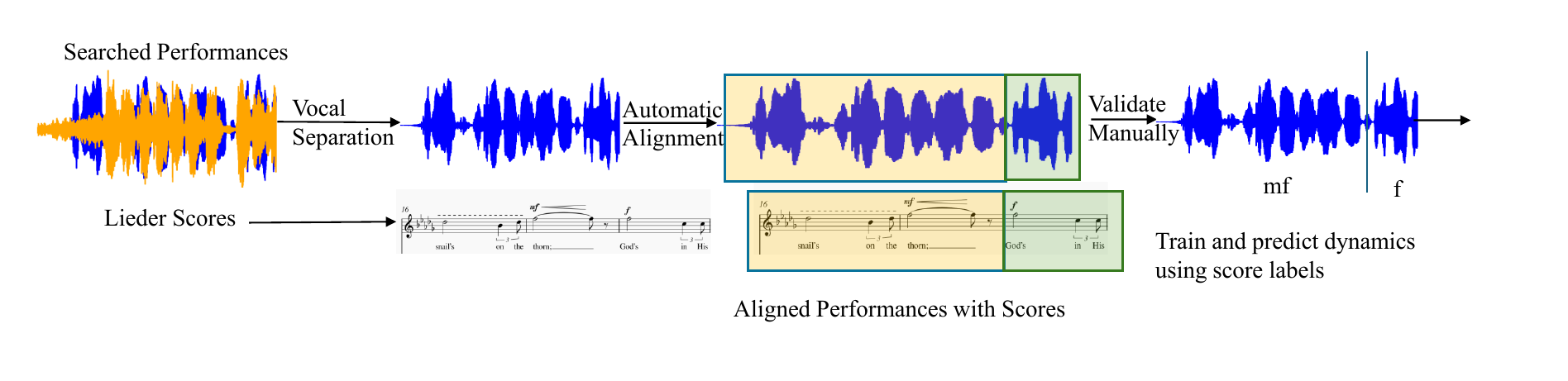}
    \caption{Data Preparation Pipeline: Corresponding to the Lieder scores from OpenScore Lieder Corpus, we apply Vocal Separation followed by Automatic Alignment. Finally, we validate the aligned score-performance data using Visualizations}
    \label{fig:overall-pipeline}
\end{figure*}
%
%
%

%
In this work, we propose to take advantage of the existing OpenScore Lieder corpus to curate a dataset of vocal performances with dynamics annotations, using state-of-the-art source separation and alignment as intermediate steps \footnote{\url{https://github.com/MTG/SingWithExpressions.git}}. Furthermore, we curate a dataset of 25 other performances of different genres annotated manually by a professional Classical vocalist to test the model. At the end, we study the relationship between score based musical dynamics to perceptually motivated audio features~\cite{friberg2014using} like log-Mel and bark-scale based features, testing the model with different analysis window-size, and genres of the test dataset. 

%
Figure \ref{fig:overall-pipeline} illustrates the overall pipeline of the task. Using the meta-data information of the repository accompanying Lieder corpus, we start with searching for corresponding performances on YouTube. Further, we apply vocal separation on the performance to get vocals. Thereafter, using state-of-art alignment techniques, we align the corresponding score with the performance. At this stage, to test the accuracy of the alignment process, we develop visualization to filter out performances with mismatched aligned scores. Using the aligned score and performance data, we train a model for estimating dynamics based markings for an unknown performance. 

The rest of this paper is structured as follows. In section 2, we cover the related works. Section 3 describes the dataset and the curation process. In section 4, we describe the experiments conducted with the curated data, followed by discussion and future work. 
\section{Related Work}
Although musical dynamics has been a topic of investigation in several studies~\cite{berndt2010modelling, bishop2014performing, elowsson2017predicting, kosta2016mapping}, especially for the case of piano~\cite{widmer2004computational, kosta2018dynamics, jeong2017note, marinelli2020musical}there remains a notable gap in research concerning standalone musical dynamics analysis for the case of singing voice, particularly from an MIR perspective. Despite this gap,
dynamics form  a fundamental aspect of analysis within the interconnected fields of singing voice synthesis~\cite{blaauw2017neural} and voice pedagogy~\cite{sundberg1994perceptual}.

In Singing Voice Synthesis (SVS) systems, dynamics play a crucial role in conveying expressive nuances~\cite{umbert2015expression}. Typically, dynamics are modelled as measures of energy in the signal~\cite{nakano2011vocalistener2, umbert2015expression, blaauw2017neural} at the frame level. However, while there exists a close correlation between energy of the signal and musical dynamics, the influence of other parameters, such as pitch and timbre~\cite{elowsson2017predicting}, remains largely unexplored. Understanding the relationship between pitch, timbre and dynamics could lead to more realistic representations of musical expression in SVS systems. 

Bous and Roebel~\cite{bous2023analysis} explore the relationship between musical dynamics and timbral characteristics of the singing voice, employing mel-spectrogram features. Their experiment involves modifying the singing voice dynamics using a neural auto-encoder to transform voice levels. Effectiveness is assessed through evaluating perceived changes in voice level in the transformed recordings. However, a significant challenge arises as there is currently no reliable labels to determine the perceived changes in musical dynamics corresponding to "voice-level" changes as proposed in the system.

Narang et al.~\cite{narang2022analysis} utilize perceptually-motivated \emph{sone} scale, comparing loudness curves of different professional renditions and student renditions for "musical dynamics" comparison following the methodology outlined by Kosta et al.~\cite{kosta2018dynamics} for comparing musical dynamics in piano. However, the study encountered limitations due to the lack of dynamics annotated datasets for evaluation. 

While there are some aspects of the research on Vocal Pedagogy~\cite{sundberg1994perceptual} that has been utilized for the case of singing voice research from an MIR perspective, for example, Phonation mode~\cite{proutskova2013breathy} dataset or VocalSet~\cite{wilkins2018vocalset} (which also contains some singing voice dynamics annotations but confined to vowel renditions), research outcomes of the vocal pedagogy remain largely unexplored by the MIR community. One direction is the role of voice source in singing voice, or how the positioning of the diaphragm affects vocal characteristics~\cite{sundberg2003research}. A study on vocal dynamics can help infer the voice source characteristics that can directly aid in vocal pedagogy.

\vspace{-2em}
\section{DataSet}\label{dataset}
Dynamics are considered to be the most commonly manipulated parameter of an expressive performance and research investigations show that professionals or experts have much better control in expressive parameters in comparison to novice performers~\cite{bishop2014performing}. Further, songs from the 19th century Romantic era of Western classical music are widely known to be rich in expressive parameters. Drawing inspiration from this notion, we curate a dataset comprising professional renditions of 19th-century songs sourced from the OpenScore Lieder corpus ~\cite{GothamJonas2022}. Notably, composers often embed numerous dynamic markings within their scores, laying a foundational framework conducive to the analysis of dynamics.
\subsection{Training Dataset Curation Process}
\subsubsection{Score Sources}
Lieder Scores is a comprehensive collection of over 1200 19th century songs encoded over several years~\cite{GothamJonas2022}. 
Within the Lieder dataset, we capitalize on two specific resources to facilitate our data curation process:
\begin{itemize}
    \item The GitHub repository of Lieder provides MSCX files along with batch-conversion script to convert to MusicXML, enabling further processing with tools such as music21~\cite{cuthbert2010music21}
    \item In the metadata section of the Lieder scores, a comprehensive compilation of composers, score names, and their respective MuseScore IDs is provided. This rich metadata serves as a valuable resource during the performance collection stage, enabling efficient querying and selection of performances.
\end{itemize}
\begin{figure}[tbp]

    \begin{subfigure}{\columnwidth}
    \centering
         \includegraphics[width=\columnwidth]{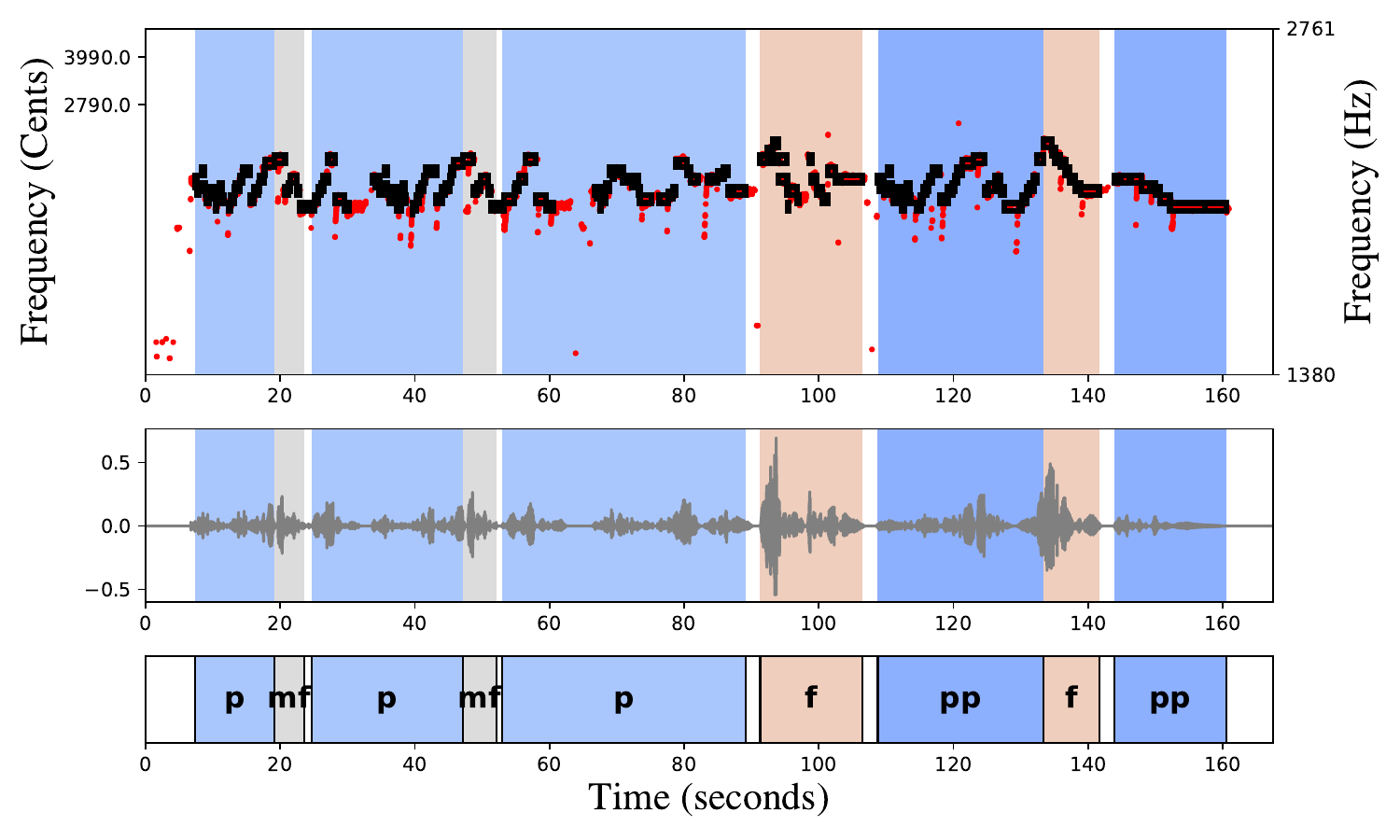}
         \caption{Accepted Performance Visualization}
         \label{fig:accepted_performance}
    \end{subfigure}
    \begin{subfigure}{\columnwidth}
    \centering
     \includegraphics[width=\columnwidth]{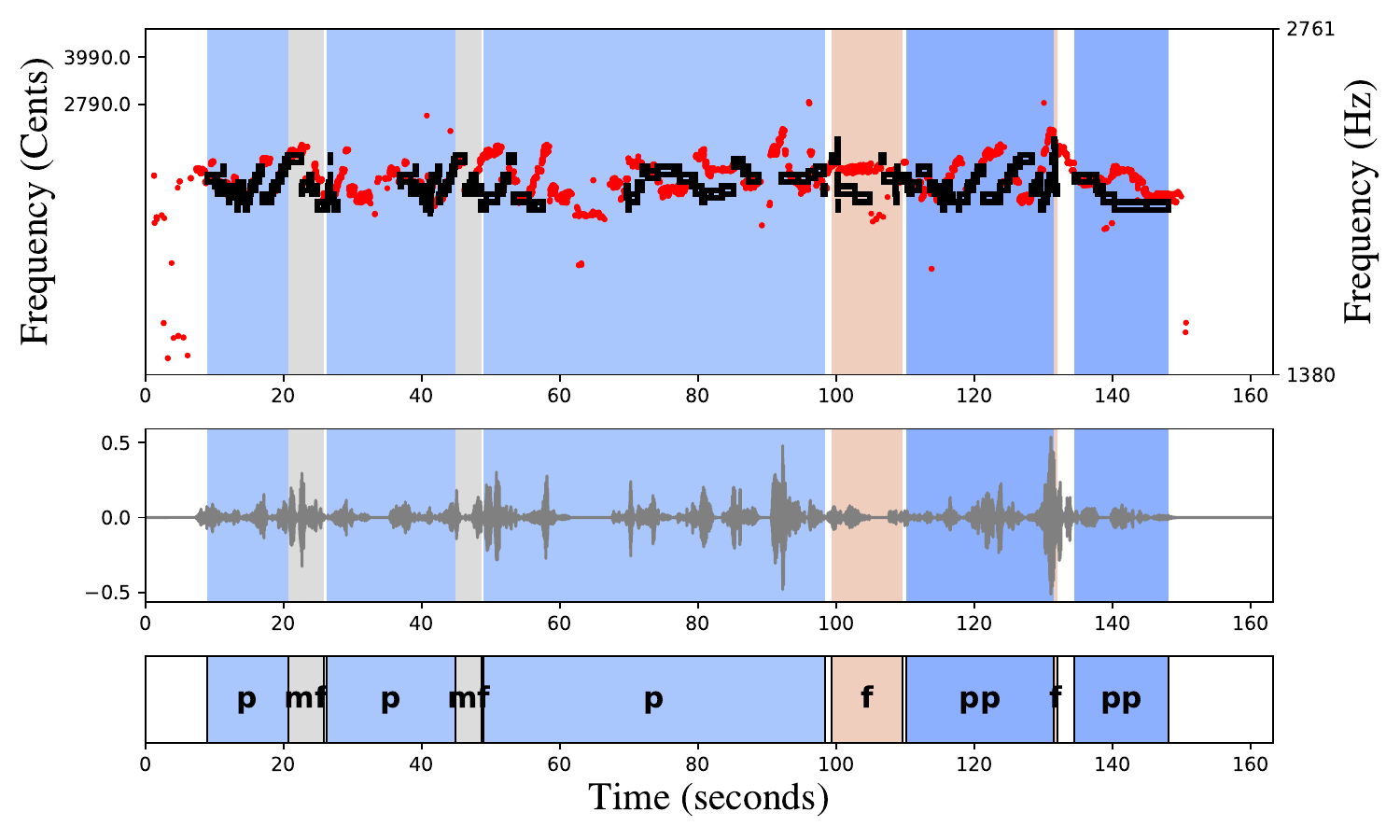}
        \caption{Rejected Performance Visualization}
         \label{fig:rejected_performance}
    \end{subfigure}
      \caption{Example visualization after automatic alignment on "The Shepherds Song" by Edward Elgar; For each sub-figure: red dots represent f0 using crepe, black dots represent note-information from the score (top), audio waveform (middle), dynamics information from the aligned score after automatic alignment (bottom)}

    \label{fig:example_brahms}
\end{figure}

\subsubsection{Filtering Criteria for Scores}
From all the batch-converted MusicXML files, we filter all scores, focusing on those with more than 3 dynamics annotations, and containing only 3 streams of score data: vocal, piano left hand and piano right hand. 

\subsubsection{Performance Sources}For the identified scores with greater than 3 dynamics markings, we search for multiple corresponding performances on YouTube using the query term obtained from the meta-data information of the scores. We curate multiple performances of similar pieces with the intention of extracting general dynamics based expressive patterns from professional singers. Our aim is to glean insights into varied interpretations, as there is no singular correct rendition of a performance that strictly adheres to the score. Subsequently, we carefully listen to each performance, specifically selecting those featuring vocals accompanied solely by piano. It is to be noted that not all composers have available performance data; thus, our selection process initially prioritizes renowned figures such as Schubert, Schumann, Brahms or Debussy, and ones with greater than 10 dynamics annotations. Once having exhaustively searched for performances of these well known composers, we proceed to search for lesser known composers following similar criteria. We automate the download process by utilizing YouTubeDL batch download to acquire the identified performances. The method yields a final list of 970 performances comprising identified composers, performances,  and their respective MusicXML score files with dynamics based annotations. 

\subsubsection{Filtering Criteria for Performances}

Following the filtration of scores and the manual curation of performance links, we advance to filtering performances suitable for the dynamics learning process. This process includes the following steps:\\
\textbf{Source Separation} Singing voices typically aren't presented in isolation. Even for solo performances, piano accompaniment is part of the performance. However, for our analysis, we require solo vocal renditions to accurately discern variations in performance dynamics. The initial step involves isolating the vocal component from the vocal-piano mix. This process, known as source separation, entails breaking a mixture into it's constituent  components, and significant research has been dedicated to separation of vocals from the mix. We use Demucs v2~\cite{defossez2021hybrid} to extract the vocals for the chosen songs. The robustness of using vocals resulting from source separation as an intermediate step was examined with the MusDB dataset~\cite{narang2021analysis}.\\ 
\textbf{Automatic Alignment} 
To ensure that our curated performances can effectively serve as the basis for dynamics analysis, it's essential to achieve a basic alignment with the scores.
Our approach to label creation draws inspiration from the methodology outlined by Tamer et al.~\cite{tamer2022violin, tamerTranscription}, who leverage Dynamic Time Warping (DTW) based music synchronization techniques~\cite{muller2021sync} for creating pseudo labels in the realm of Violin transcription.
Additionally, the concept of utilizing audio-to-score alignment as a pre-processing step for curating datasets in a semi-automatic manner for musicological endeavours was introduced in works by Weiss et al.~\cite{WeissZAGKVM20_WinterreiseDataset_ACM-JOCCH}, with a focus on the curation of Schubert's Winterreise dataset. While the works by Weiss et al. utilize MIDI-to-score alignment, we have chosen to conduct the alignment using musicXML scores. This decision stems from the fact that dynamics information such as \textit{piano, forte, crescendo, and diminuendo} can be less reliable in the process of MIDI conversion.\\
\textbf{Manual Filtering using Visualizations} The alignment stage yields a score with time information mapped to the corresponding performance files. Subsequently, we develop a visualization process utilizing fundamental frequency (f0) data extracted from performance files using CREPE~\cite{kim2018crepe} to validate the alignment between time-aligned performance and score files. Figure~\ref{fig:example_brahms} showcases a sample visualization from the dataset. Figure \ref{fig:accepted_performance} illustrates a performance that was accepted, and Figure \ref{fig:rejected_performance} depicts a performance that we manually excluded during the selection process. The performance in Figure \ref{fig:rejected_performance} was rejected because f0 curve from crepe (red dots) do not align with note-information from score (black rectangles) after automatic alignment, and hence the final labels lose reliability. The end result of this step is a comprehensive dataset of 509 performances for 163 aligned score files, which can be used to extract precise note-level expressive information from the score using tools like music21~\cite{cuthbert2010music21}.
\subsubsection{Dynamics based Labels Extraction from Aligned Score Files} 
The aligned score files consist of all score-based information crucial for dynamics prediction. In this stage, we process the musical dynamics labels extracted using music21. Our approach adheres to the following principle:  consecutive notes in the aligned audio are assumed to maintain similar dynamics unless there is a change in dynamics annotation in the score. When encountering labels like \textit{sfz} or \textit{sf} for a note, the value of the label of the consecutive note is assigned to be the dynamic value of the note preceding \textit{sf} or related categories. This process results in a note-level mapping of 13 musical dynamics categories: \textit{pppp, ppp, pp, p, mp, mf, f, ff, fff, ffff, sf, crescendo, diminuendo} directly extracted from the score. It is to be noted that we consolidate accent related categories, such as \textit{sf}, \textit{sfz} into a single category. Additionally, while we focus on musical dynamics for our task, the aligned score-performance data holds potential for various other Music Information Retrieval (MIR) tasks related to singing voice, including transcription, synthesis, or pedagogy.

\begin{figure}
    \begin{subfigure}{0.48\columnwidth}
    \centering
        \includegraphics[width=\textwidth]{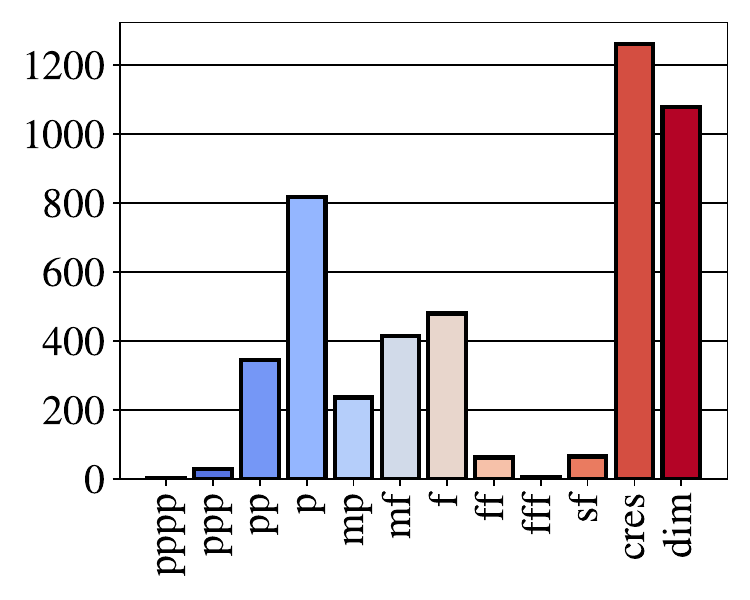}
        \caption{Dynamics Distribution Count for 163 Training Data Score Files}
    \end{subfigure}
  \hfill
    \begin{subfigure}{0.48\columnwidth}
    \centering
        \includegraphics[width=\textwidth]{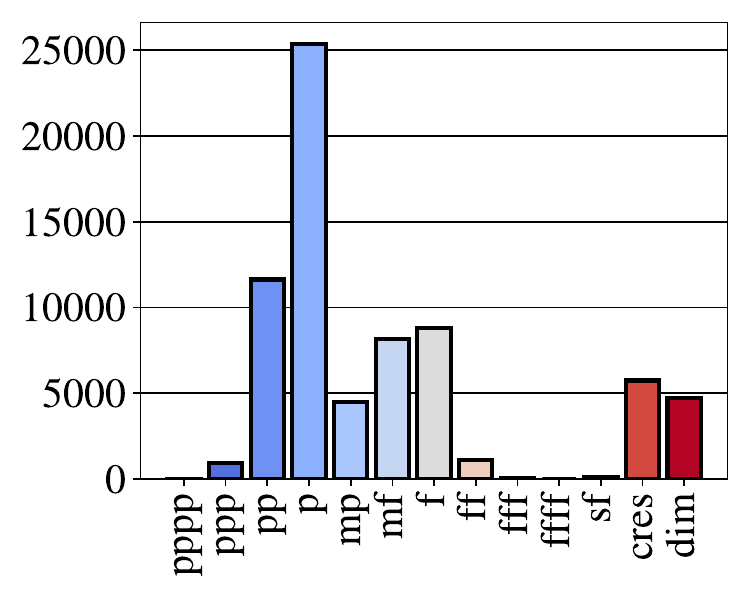}
        \caption{Distribution by Time in Filtered Training Audio Performances (seconds)}
        \label{fig:sub2}
    \end{subfigure}
    
    \begin{subfigure}{0.48\columnwidth}
     \centering
        \includegraphics[width=\textwidth]{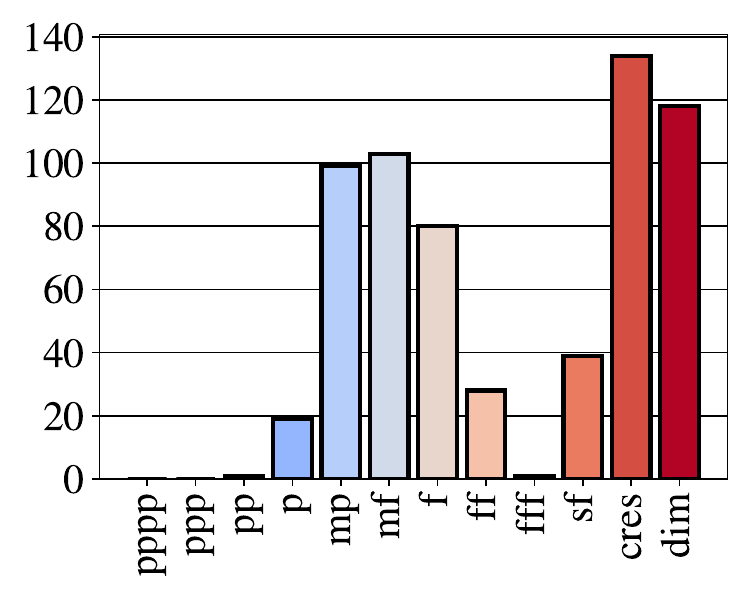}
        \caption{Dynamics Distribution Count for 25 Testing Data Score Files}
        \label{fig:sub3}
    \end{subfigure}%
    \hfill
    \begin{subfigure}{0.48\columnwidth}
     \centering
        \includegraphics[width=\textwidth]{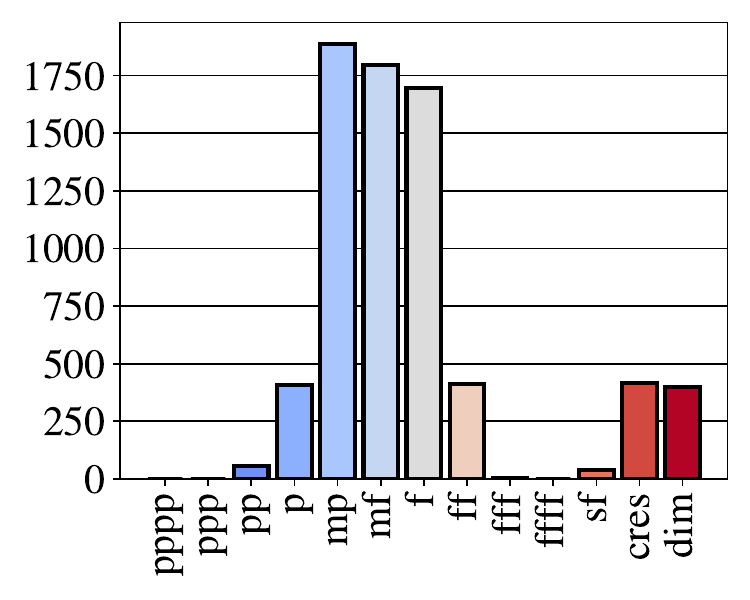}
        \caption{Distribution by Time in Filtered Testing Data Audio Performances (seconds)}
        \label{fig:sub4}
    \end{subfigure}
    \caption{Dynamics Distribution across Train and Test Performances}
    \label{fig:dynamics_distribution_of_score_files}
\end{figure}
\subsection{Test Dataset Curation Process}
For testing, we curated performances from a diverse selection of genres, ranging from operatic pop to theatre, R\&B, or jazz, which lie outside the typical classical music domain. 
We collaborated with a Classical Vocalist, possessing over a decade of experience, to identify artists renowned for their wide vocal range. Once identified, we created reference scores for selected performances by these professional artists. The distribution of the genres in the selected pieces is as follows: pop(13), rock(12), jazz(3), soul(5), R\&B(5), theatre(2) and other miscellaneous genres(5) including categories such as "post-disco", "acoustic" or "progressive rock", amongst others. 


\subsubsection{Annotation Methodology} 
This section details the annotation methodology for dynamics-related markings of selected pieces as outlined by the musician:
In the first listening, the piece's starting dynamic value is determined according to the dynamic markings such as: \textit{pp, p, mp, mf, f, ff}, creating a reference point for each piece. This phase captures the most prominent features, recognizing that notation conveys more than mere amplitude. 
Subsequent listenings entail adding details, both in terms of dynamics and articulation of the text and musical phrases. Increased attention reveals additional layers of variation, often unnoticed during the first listening. 
In the third listening, decisions are made based on unification criteria. If different notations were used for the same musical effect in similar portions of the piece (e.g., different verses), the notation that best represents the musical intent is selected and unified with the rest. Rarely, genuine differences may exist between similar sections, in which case they are left distinct. 
In the final listening, no further notations are added. Instead, a mental musical reading of the entire work, from beginning to end, is undertaken. This involves elaborating on the interpretation following the written notations while simultaneously comparing it with the rendition produced by the artist.


    
\begin{table*} 
\centering
\caption{Results with Mel and Bark Features. Temporal resolution refers to the final feature rate after downsampling.}
\label{tab:results}
    \begin{tabular}{lllccc}\toprule   
    Seq Length & Temporal Resolution & Perceptual Feature & Acc & Acc($\pm1$)  & Acc($\pm2$) 
    \tabularnewline\midrule
    4096 & 17.4 ms & log-Mel & 6.95 & 38.46 & 63.02
    \tabularnewline
    10000 & 29 ms & log-Mel & 11.35 & 42.55 & 68.38
    \tabularnewline
    4096 & 16 ms & Bark & 20.44 & 59.17 & 82.24 
    \tabularnewline
    10000 & 30 ms & Bark & 20.96 & 60.71 & 84.78
    \tabularnewline\bottomrule
    \end{tabular}
\end{table*}
\subsubsection{Processing Methodology for Test Dataset}
The processing methodology followed for the test dataset is similar to that of the training dataset, i.e., we apply source separation followed by automatic alignment to fetch the annotated labels using curated reference scores and performances. 
\subsection{Dataset Statistics}

\textbf{Audio Statistics}: The total duration of all the performances for the training dataset is 25.91 hours. The total duration of test files is 1.614 hours. The distribution of the labels as identified in dataset section \ref{dataset} is illustrated in Figure \ref{fig:dynamics_distribution_of_score_files}. 
We observe that Lieder scores follow a relatively uniform distribution of dynamics with large number of dynamics annotations centered on a `\textit{piano}'. And for the test dataset, the distribution curve is largely gaussian with majority of the distribution centered around  \textit{mp} and \textit{mf}, which is not surprising considering the nature of pop music and mixing and mastering effects added to the final renditions.\\
\textbf{Performance Count Per Piece}: Although a single performer can deviate from the annotated score dynamics, having multiple performers per piece can help the model learn the general patterns closer to composer's intention. To leverage this effect, we collect performances with an average count of 3.12 performances per piece (std: 2.13), with a maximum of 12 performances for a piece by Robert Schumann. The average performance duration was observed to be 9.54 minutes (std: 9.36 minutes), with a maximum of 74.27 minutes for a piece by Franz Schubert and a minimum of 1.01 minutes for a piece by Peter Warlock. 

\section{Experiments and Results}
For the experiments outlined in this section, we utilize the curated dataset of Classical vocal performances for training and the dataset created in collaboration with the Classical vocalist for testing. We convert the note-level dynamics labels  spanning from \textit{pianissississimo (pppp)} to\textit{ fortissississimo (ffff)} into framewise labels encompassing $10$ dynamics classes, and train and test our models for estimating the frame-wise dynamics. Thus, we consider dynamics estimation as a 10-class classification problem operating at the granularity of individual frames.


\noindent \textbf{Input Representations}: 
For model inputs, we consider two perceptually-motivated loudness features that are extracted after isolating the vocal tracks using DemucsV2\cite{defossez2021hybrid}. As our first input representation, we consider log-Mel features, which are commonly used in many audio and music processing tasks. These features are extracted using the librosa~\cite{mcfee2015librosa} library from audio sampled at 44.1 kHz using a hop size of 5.8 ms. As our second representation, we consider the specific loudness in Bark critical bands, which was previously studied in the context of piano dynamics \cite{kosta2018dynamics} and singing voice loudness analysis ~\cite{narang2022analysis}. The 240 dimensional Bark features are extracted using the MoSQITo library~\cite{green_forge_coop_2024_10629475, san2021mosqito}, following the Zwicker loudness calculation method for time-varying signals \cite{zwicker1991program} as specified in the ISO.532-1:2017 standard. The extraction process adheres to the default settings of a 48 kHz audio sampling rate and a 2 ms hop size.

Alongside these different input representations, we also study the effect of input sequence length and rate. To that end, we experiment with sequence lengths of $4096$ and $10000$. Since the original input representations have different temporal resolutions, we employ various downsampling rates to ensure that the models receive comparable feature rates during analysis. In our study with short context ($4096$ frames) dynamics modeling, we downsample the Bark features by $8$ to operate at $16$ ms, and downsample the log-Mel features by $3$ to operate at $17.4$ ms. For modeling dynamics detection using longer contexts ($10000$ frames), we downsample the log-Mel features by $5$ to achieve a temporal resolution of $29$ ms, and downsample the Bark features by $15$ to achieve a comparable resolution of $30$ ms.

\noindent \textbf{Model Architecture and Training}: For the frame-level estimation of dynamics, we employ a multi-scale Convolutional Neural Network (CNN) with self-attention\footnote{Based on the modified version of \url{https://github.com/LiDCC/GuzhengTech99/blob/main/function/model.py}}~\cite{li2023frame_ipt}, originally introduced for the closely related task of frame-wise playing technique detection. In our implementation, the network receives input features with a fixed sequence length and outputs probabilities for 10 dynamics classes, with the class having the highest probability taken as the estimate. During training, we utilize the Adam optimizer with a learning rate of 0.002, aiming to reduce the Cross Entropy loss between the predicted dynamics classes and the aligned dynamics labels. We report our results on training the same network for different input representations, sequence lengths, and feature rates.

\noindent \textbf{Metrics}: One big challenge in the experimentation with musical dynamics is the subjectivity and relativity in its evaluation. For instance, one piece may span dynamics ranging from \textit{pp} to \textit{f}, and another piece may span dynamics ranging from \textit{p} to \textit{ff}. However, the measured loudness values of performances derived from both music pieces might be similar, as both sets of labels indicate a transition from relatively "soft" to "loud" dynamics. Therefore, the mapping between perceived performed dynamics and labeled musical dynamics may not be absolute. To address this challenge, we present the results in terms of exact match (\text{Acc}), relaxed accuracy 1($\text{Acc}\pm1$) , and relaxed accuracy 2($\text{Acc}\pm2$). Relaxed accuracy denotes that estimates are not penalized for a mismatch of 1 or 2 classes, respectively.\\

\vspace{-3em}
\subsection{Results}
\begin{figure}[tbp]
    \centering
    \includegraphics[width=\columnwidth]{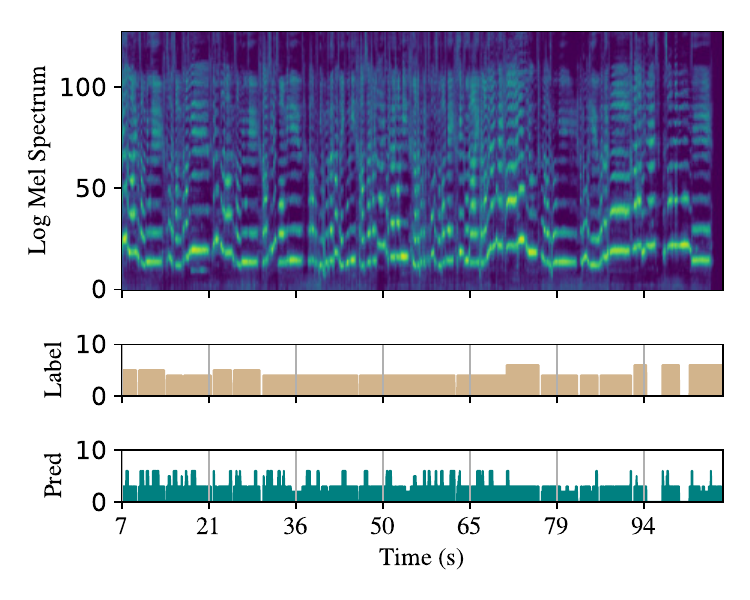}
    \caption{Model input and outputs for the log-Mel spectrum features. log-Mel-spectrogram (top), annotated labels by musician(middle), model estimates (bottom)}
    \label{fig:log-mel}
\end{figure}

\begin{figure}[tbp]
    \centerline{
        \includegraphics[width=0.99\columnwidth]{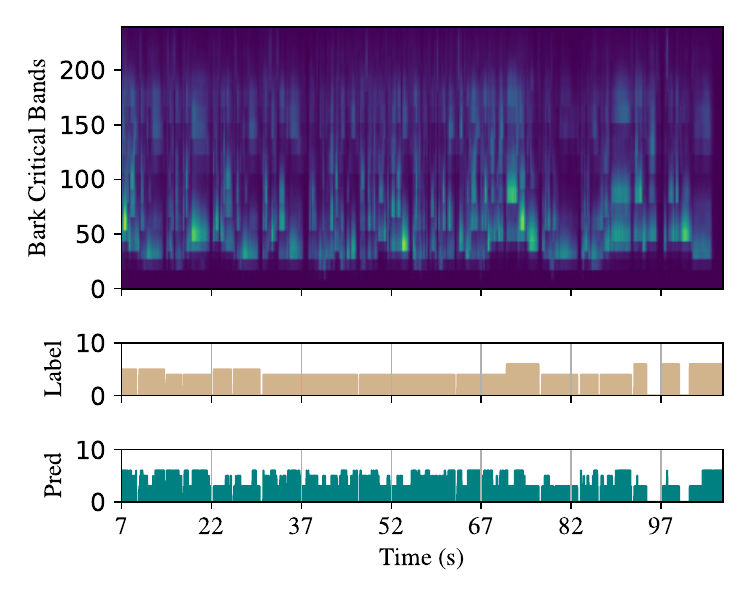}}
         \caption{{Model input and outputs for the bark based features: bark-critical-bands (top), annotated labels by musician (middle), model estimates (bottom)}}
    \label{fig:bark-based}
\end{figure}

The results are summarised in Table \ref{tab:results}. Despite the subjectivity of the task, we observe that the most confusion lie within the $\pm1$ or $\pm2$ range with significantly higher relaxed accuracies. Furthermore, we see that bark-based features outperform log-Mel features for the task. The highest relaxed accuracy $\pm2$ is achieved with bark-based features, indicating the models ability to differentiate between upper and lower bounds of dynamics. For example, a fortissimo is not classified to be a piano in almost 85\% of the cases. An example prediction using log-Mel features and bark-based features for a theatre song "sound of music" is presented in Figures \ref{fig:log-mel} and \ref{fig:bark-based} respectively. 

The effect of larger and smaller temporal contexts can also be seen in Table \ref{tab:results}. Providing larger temporal contexts results in better performance for dynamics estimation. This effect is more prominent for log-Mel features compared to the Bark features. We found that the best performing model is the one with the entire song frames included in the context window i.e., the sequence length is long enough to encapsulate the whole song.



\section{Discussion}
One of the primary challenges in predicting musical dynamics lies in the fact that performance information is available through recordings, which is a result of mixing and mastering. Consequently, the loudness information captured in recordings may diverge from performers original intentions. However, we contend that despite the influence of mixing and mastering, it is possible for musicians as well as non-musicians to infer whether a performer is singing softly, loudly or even shouting independent of raw loudness levels. Our approach leverages perceptually motivated features that encapsulate timbral characteristics, which have the potential to enhance musical dynamics estimation while remaining agnostic to variations in loudness levels.

While the labels are created semi-automatically, there are potential discrepancies due to performers not adhering strictly to the score or editors creating alternative versions of the score different from the one curated in the dataset. 

Additionally, we have framed the dynamics estimation at an absolute level, with the expectation that the model will learn the variations in relative markings given a large amount of data. However, musical dynamics at any given time in a performance depend on the context rather than the absolute value of measured loudness~\cite{nakamura1987communication}. Additionally, addressing class imbalance remains a significant challenge. 

On software front, while MuseScore offers extensive annotation capabilities, some categories cannot be accurately modeled. To mitigate this, musicians often use note-level "TextExpressions" in MuseScore to add additional information. During our experimentation, we encountered terms like "sempre piano," "poco dolce," and "calando" that musicians add to the score. While we were able to mitigate challenges with some labels, achieving comprehensive coverage requires further collaboration with vocalists to refine the target labels.

\section{Conclusion and Future Work}

We've developed a methodology for large-scale dataset curation focused on singing voice. The semi-automatically curated dataset serves as a valuable resource for tasks such as transcription, expression analysis, synthesis, and vocal pedagogy. It currently includes 509 performances aligned with 163 score files from 25 composers. Using this dataset, we trained a CNN with multi-head attention for dynamics prediction and found that bark-scale-based features outperform log-Mel features. To test the model, we curated score-performance dataset manually in collaboration with a Classical vocalist. Future work involves integrating pitch features with loudness features to enhance prediction accuracy, improving the model to address class imbalance, and expanding the dataset to include more composers.

\section{Acknowledgments}
We would like to thank Ajay Srinivasamurthy for his invaluable feedback. IA y Música: Cátedra en Inteligencia Artificial y Música" (TSI-100929-2023-1), funded by the Secretaría de Estado de Digitalización e Inteligencia Artificial, and the European Union-Next Generation EU, under the program Cátedras ENIA 2022 para la creación de cátedras universidad-empresa en IA. 



\bibliography{ISMIR2024_template}

%
%
%
%
%

\end{document}